\documentclass{article}

\PassOptionsToPackage{numbers}{natbib}
\usepackage[preprint]{neurips_2025}
\usepackage{xspace}
\usepackage{amsmath}
\usepackage{makecell}

\usepackage[utf8]{inputenc} 
\usepackage[T1]{fontenc}    
\usepackage{hyperref}       
\usepackage{url}            
\usepackage{booktabs}       
\usepackage{amsfonts}       
\usepackage{nicefrac}       
\usepackage{microtype}      
\usepackage{xcolor}         
\usepackage{graphicx}
\usepackage{wrapfig}
\usepackage[inline]{enumitem}
\usepackage{subcaption}

\title{Visuo-Acoustic Hand Pose and Contact Estimation}

\author{%
  Yuemin Mao $^{1}$$^{*}$ \quad 
  Uksang Yoo$^{1}$$^{*}$\quad
  Yunchao Yao$^{1}$ \quad
  Shahram Najam Syed$^{1}$ \And
  Luca Bondi$^{2}$ \quad
  Jonathan Francis $^{1,2}$ \quad
  Jean Oh$^{1}$ \quad
  Jeffrey Ichnowski$^{1}$
}
\newcommand{\algname}{VibeMesh\xspace}

\begin{document}

\maketitle
\begin{center}
    \vspace{-1.5em}
    \textsuperscript{*}Equal contribution. \\[1ex]
    \textsuperscript{1}Robotics Institute, Carnegie Mellon University, Pittsburgh, USA \\[1ex]
    \textsuperscript{2}Bosch Center for Artificial Intelligence, Pittsburgh, USA
    \vspace{1.5em}
\end{center}

\begin{abstract}
Accurately estimating hand pose and hand-object contact events is essential for robot data-collection, immersive virtual environments, and biomechanical analysis, yet remains challenging due to visual occlusion, subtle contact cues, limitations in vision-only sensing, and the lack of accessible and flexible tactile sensing. We therefore introduce \textbf{\algname}, a novel wearable system that fuses vision with active acoustic sensing for dense, per-vertex hand contact and pose estimation. \algname integrates a bone-conduction speaker and sparse piezoelectric microphones, distributed on a human hand, emitting structured acoustic signals and capturing their propagation to infer changes induced by contact. To interpret these cross-modal signals, we propose a graph-based attention network that processes synchronized audio spectra and RGB-D-derived hand meshes to predict contact with high spatial resolution. We contribute: (i) a lightweight, non-intrusive visuo-acoustic sensing platform; (ii) a cross-modal graph network for joint pose and contact inference; (iii) a dataset of synchronized RGB-D, acoustic, and ground-truth contact annotations across diverse manipulation scenarios; and (iv) empirical results showing that \algname outperforms vision-only baselines in accuracy and robustness, particularly in occluded or static-contact settings.
\end{abstract}

\section{Introduction}

Accurately estimating human hand pose \emph{and} contact is critical for robot teleoperation~\cite{qin2023anyteleop, zick2024teleoperation, cheng2024open, yin2025dexteritygen}, virtual reality~\cite{wu2020hand, ahmad2019hand}, and biomechanical analysis~\cite{coupier2016novel, salchow2019tangible, rath2011hand}. In all of these settings, knowing \textit{when} and \textit{where} the hand touches the environment—together with its configuration—enables reasoning about task phases, distinguishing exploration from manipulation, and inferring force dynamics. Unfortunately, real‑world contact sensing is hard: occlusions, limited sensor viewpoints, and subtle touch events routinely confound purely visual approaches.

Vision‑based methods typically estimate contact indirectly, pairing RGB or depth observations with strong priors on object geometry and canonical hand poses~\cite{fan2024hold, brahmbhatt2020contactpose, ye2023diffusion, swamy2024showme}. Model‑based fitting can help~\cite{wang2023deepsimho, wang2023interacting, kuang2024learning}, but still fails under poor lighting and suffers from ambiguities due to occlusion. Direct tactile solutions, for example, capacitive and piezoelectric gloves~\cite{luo2024adaptive, hubble2019sensing} or full‑body suits~\cite{fujimori2009wearable}, offer better signal fidelity at the cost of bulk, expense, and limited practicality. Meanwhile, promising advances in wearable acoustics~\cite{yu2024ring} and cross‑modal learning~\cite{tatiya2024mosaic} have yet to be exploited for dense hand-object contact estimation.

\begin{figure}[t]
\centering
\includegraphics[width=1.0\columnwidth]{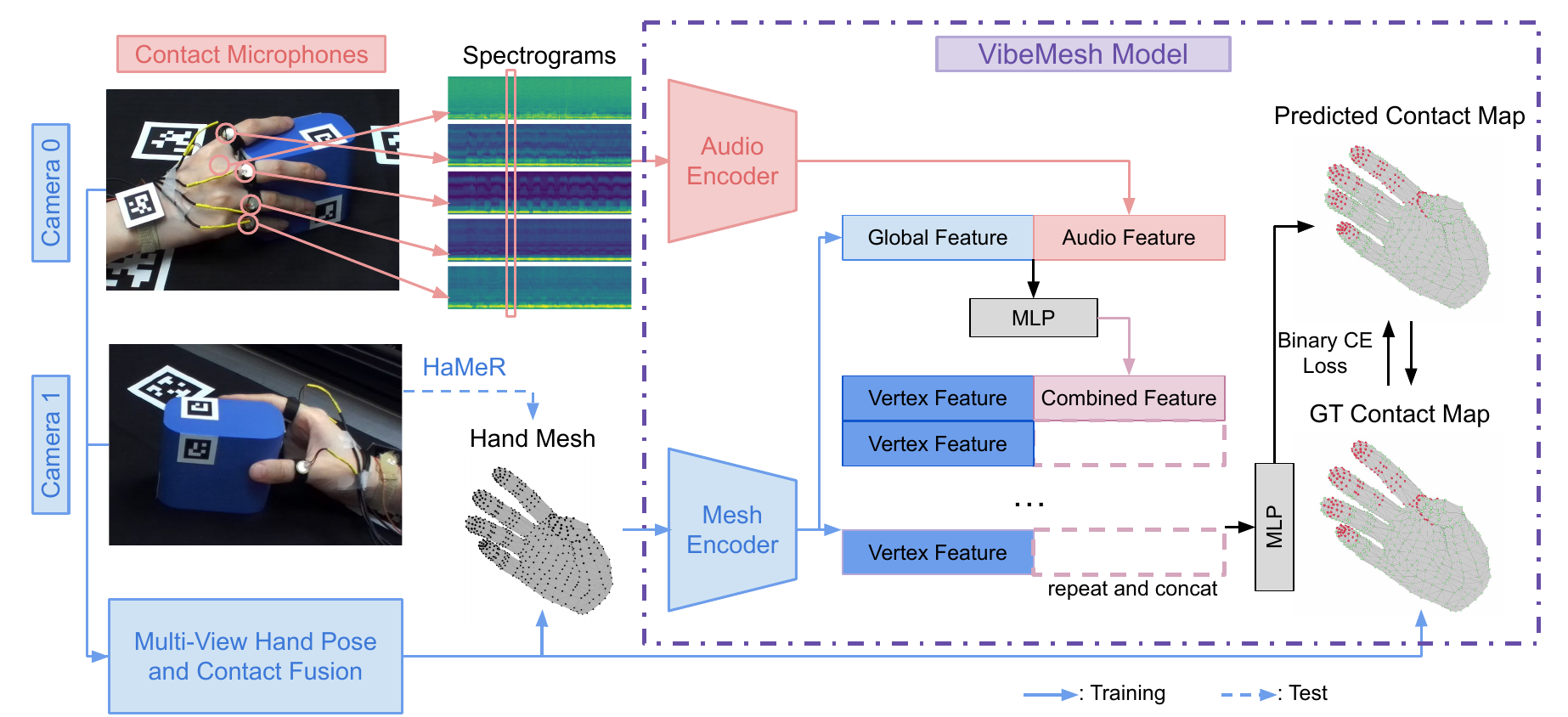}
\caption{\textbf{Overview of \algname.} Our visuo-acoustic contact estimation architecture predicts per-vertex contact by integrating audio embeddings with hand-mesh features. During training, it leverages audio with hand meshes and contact annotations jointly reconstructed from two synchronized RGB-D camera streams; at inference, it operates on audio and a hand mesh reconstructed from a single, partially occluded, RGB view, demonstrating robustness under visual ambiguity.}
\label{figure:overview}
\end{figure}

We propose bridging this gap with a \emph{visuo‑acoustic} sensing approach that delivers joint pose- and contact-estimation. A lightweight bone‑conduction speaker, mounted on the wrist, emits a signal that consists of a wide range of acoustic frequencies (a process called `broadband probing'), whose propagation behavior changes whenever the hand changes configuration or touches an object. A sparse array of piezoelectric contact microphones on the fingers records these shifts, providing rich, self‑generated cues that remain informative even when the hand is static or visually occluded. To interpret them, we propose \textbf{\algname}: a cross‑modal graph‑attention network that fuses spectral audio features with a mesh‑based MANO~\cite{MANO2017} hand representation to predict per‑vertex contact labels from synchronized audio and RGB input (Fig.~\ref{figure:overview}).

This paper contributes: 
\begin{enumerate*}[label=(\textit{\roman*})]
    \item a wearable visuo‑acoustic platform for contact‑aware hand tracking;
    \item \algname, a cross‑modal graph‑attention architecture that fuses acoustic and visual cues for dense contact prediction; 
    \item a dataset of time‑aligned RGB‑D, audio, and ground‑truth contact annotations across diverse grasps; and
    \item thorough evaluation across users and objects demonstrating improved accuracy and robustness under occlusions, diverse object properties, and static‑contact scenarios compared against state‑of‑the‑art baselines.
\end{enumerate*}

\section{Related Works}

\subsection{Visual Hand Pose and Contact Estimation}
Researchers have extensively studied vision-based hand pose and contact estimation motivated by its application to VR/AR/XR~\cite{wu2020hand, ahmad2019hand,marchand2015pose, bun2022hand}, robotics~\cite{qin2023anyteleop, zick2024teleoperation, cheng2024open, yin2025dexteritygen, niu2025human2locoman}, and kinesiology~\cite{airo2016vision, visee2020effective}. The ubiquity of cameras in wearable devices and everyday environments makes vision a convenient and accessible sensing modality~\cite{zheng2023deep}. 

Classical approaches primarily relied on template matching and silhouette-based techniques~\cite{erol2007vision, rosales20013d}. More recent methods leverage large-scale synthetic~\cite{li2023renderih, cao2021reconstructing} or real-world datasets~\cite{xiang2019monocular, simon2017hand, moon2018v2v, chao2021dexycb, hampali2020honnotate, zimmermann2019freihand} and expressive models to regress 2D keypoints or full 3D meshes from monocular RGB inputs~\cite{zimmermann2017learning, fan2024hold}. Transformer-based architectures~\cite{jiang2023a2j, hampali2022keypoint, wen2023hierarchical, pavlakos2024reconstructing} and physics-informed pipelines~\cite{wang2023deepsimho, tzionas2016capturing} have further improved pose estimation accuracy and robustness.

However, vision-only methods remain fundamentally limited by self-occlusion and inherent visual ambiguities, particularly during hand-object interactions~\cite{fan2021learning}. To address these limitations, \algname incorporates audio as an additional sensing modality. By exploiting the rich temporal and spectral structure of contact-induced sounds, which reflect changes in the lumped acoustic properties of the hand-object system, \algname learns to disambiguate visually similar hand poses and to localize contact more precisely than vision-only approaches.

\subsection{Multi-Modal Human Pose Tracking}

To address the inherent limitations of vision-only systems, recent work has explored incorporating additional sensing modalities such as tactile~\cite{park2024stretchable, tashakori2024capturing, de2022human}, EMG~\cite{seo2024posture, yoshikawa2007hand, salter2024emg2pose, liu2021neuropose}, and audio~\cite{lee2024echowrist, wang2025ram, wang2024towards, shibata2023listening}. These modalities provide complementary signals that capture muscle activity, contact forces, and acoustic transients associated with physical interactions, enabling robust pose and contact estimation in the presence of occlusion or visual ambiguity.

Tactile and stretchable sensors offer direct measurements of contact and deformation but often require expensive, form-fitting gloves or embedded skins that can restrict natural user motion~\cite{belcamino2024systematic}. EMG-based approaches infer pose from muscle activation, yet are sensitive to electrode placement and prone to signal drift or degradation due to user fatigue~\cite{salter2024emg2pose, sun2022application}. Additionally, tactile sensor-embedded gloves generally suffer from low spatial resolution, often segmenting the hand into a few contact patch regions~\cite{seo2024posture}. In contrast, audio sensing offers an indirect but non-intrusive alternative~\cite{yu2024ring}, with sound carrying rich temporal and spectral cues indicative of contact timing, material properties, and dynamic interactions—even when visual signals are occluded. By leveraging synchronized audio-visual signals, \algname learns to disambiguate occluded hand poses and accurately localizes contact events, while minimizing reliance on wearable or invasive instrumentation.

\section{Methods}

\algname consists of three stages: data collection, vision-based ground-truth contact labeling, and model training. In the data collection stage, we capture multi-model recordings by synchronizing two RGB-D cameras with a set of wearable acoustic sensors. Next, we merge the video streams to produce ground-truth 3D hand poses and contact annotations. For model training, we learn a network that fuses mesh features from a graph-attention encoder with CNN-extracted audio embeddings, optimizing contact prediction via binary cross-entropy against the ground-truth contact labels.

\subsection{Visuo-Acoustic Hand-Object Interaction Data Collection}
We develop a system that integrates wearable acoustic sensors with RGB-D cameras to capture multi-model data for hand-object interactions. We collect a dataset from multiple users performing a wide range of grasping conditions.


\begin{figure}[t]
  \centering
  \begin{subfigure}[t]{0.33\textwidth}
    \centering
    \includegraphics[width=\linewidth]{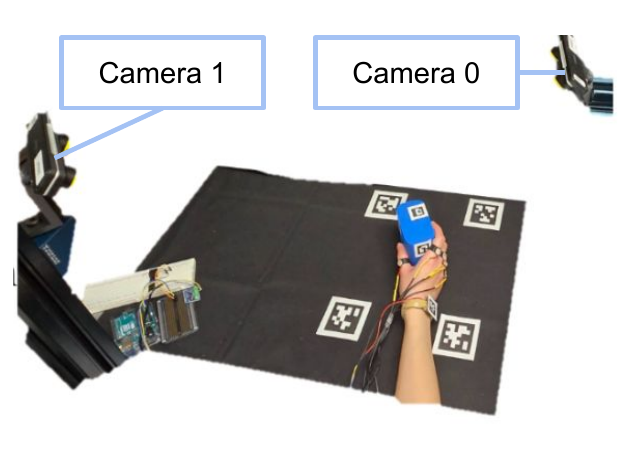}
    \caption{Data collection workspace.}
    \label{figure:data_collection_ws}
  \end{subfigure}%
  \hfill%
  \begin{subfigure}[t]{0.33\textwidth}
    \centering
    \includegraphics[width=\linewidth]{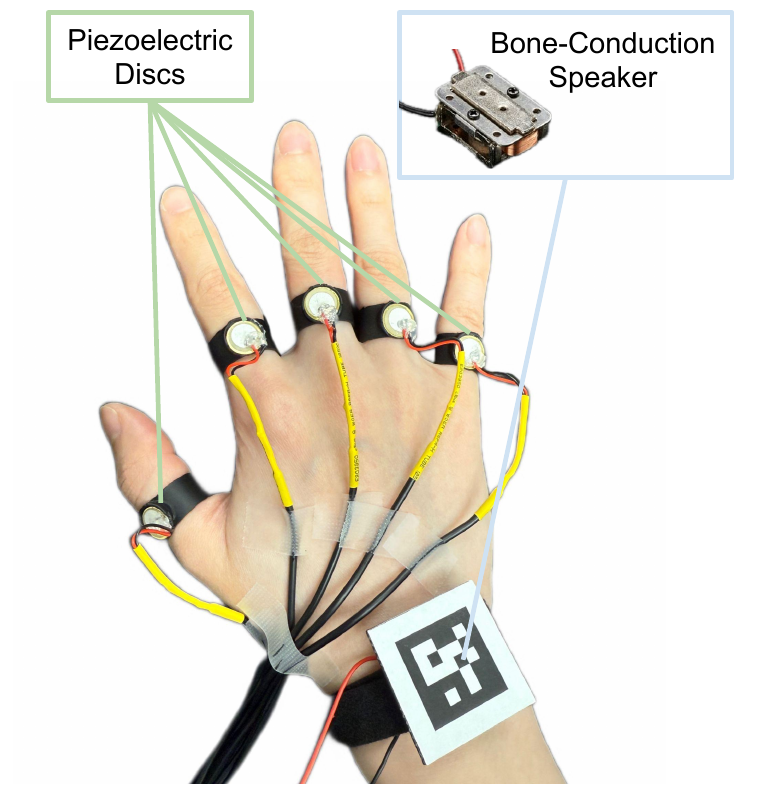}
    \caption{Wearable microphone rings.}
    \label{figure:wearable}
  \end{subfigure}%
  \hfill%
  \begin{subfigure}[t]{0.33\textwidth}
    \centering
    \includegraphics[width=\linewidth]{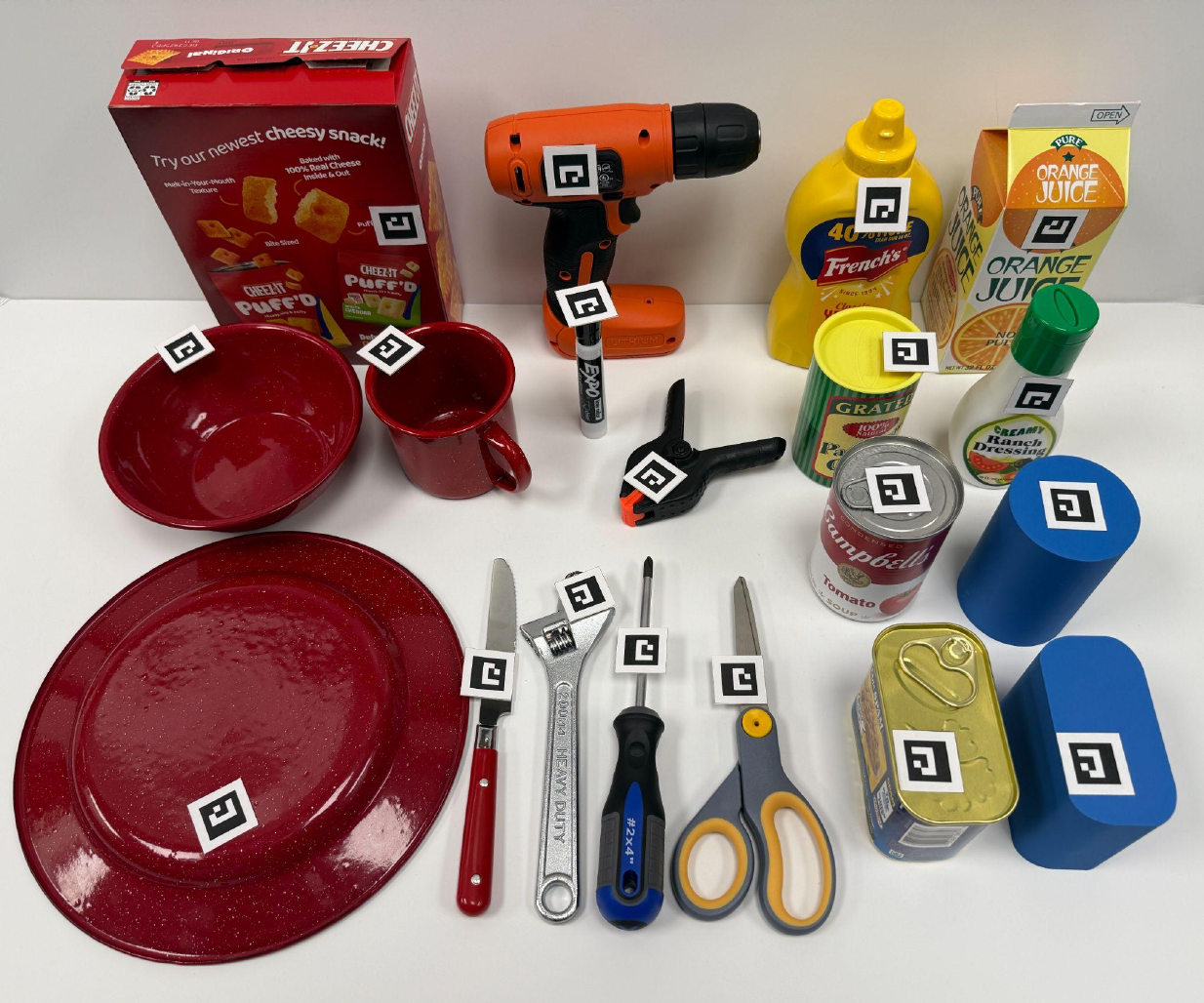}
    \caption{Data collection objects.}
    \label{figure:object}
  \end{subfigure}
  \caption{\textbf{\algname data collection setup.} (a) Our multi-view capture station with two calibrated ZED Mini RGB-D cameras positioned at complementary angles to minimize occlusion during hand-object interaction. (b) Custom-designed piezoelectric microphone rings worn on each finger, with the bone-conduction speaker mounted on the wrist. Each ring contains a 10\,mm piezoelectric disc sensor connected to a Maono USB sound card. (c) The diverse set of 19 objects used in our experiments includes items from YCB and HOPE datasets with varied geometries, materials, and affordances to evaluate across different manipulation scenarios.}
\end{figure}

\paragraph{Wearable Microphones and Speaker}
For active acoustic sensing, we use an Adafruit bone-conduction speaker to generate white noise as the signal, paired with five contact microphones (10\,mm piezoelectric discs) for signal acquisition at a sampling rate of 44.1\,kHz. Each contact microphone is interfaced with a Maono USB sound card and mounted onto a 3D printed PLA ring, sized appropriately for specific user fit. During data collection, participants wear the rings on their fingers, with the speaker secured to their wrist using an elastic band (Fig.~\ref{figure:wearable}).

\paragraph{RGB-D Cameras} To obtain high-quality ground-truth labels for hand pose and contact, we use two ZED Mini stereo RGB-D cameras positioned at complementary viewing angles (Fig.~\ref{figure:data_collection_ws}) to reduce occlusions. Both cameras record synchronized data at a resolution of 1080p and a frame rate of 30\,fps, and operate with Zed AI disparity estimation to enhance depth map resolution and accuracy.

\paragraph{Data Collection Procedure}
Five right-handed participants (3 male and 2 female) with varied hand and finger dimensions took part in data collection under IRB approval. We selected 19 objects with high-quality meshes varying in geometry, mass, and material (Fig.~\ref{figure:object}): 14 from the YCB dataset~\cite{calli2017ycb}, 3 from the HOPE dataset~\cite{tyree2022hope}, and 2 PLA-printed replicas matching 2 YCB objects. Each participant completed five 60-second sessions per object, during which they repeatedly grasped and released the item and were instructed to creatively explore different grasp poses to ensure a diversity of hand poses and contact conditions.


\subsection{Vision-Based Hand Pose and Contact Estimation}
\begin{figure}[t]
\centering
 \includegraphics[width=1.0\columnwidth]{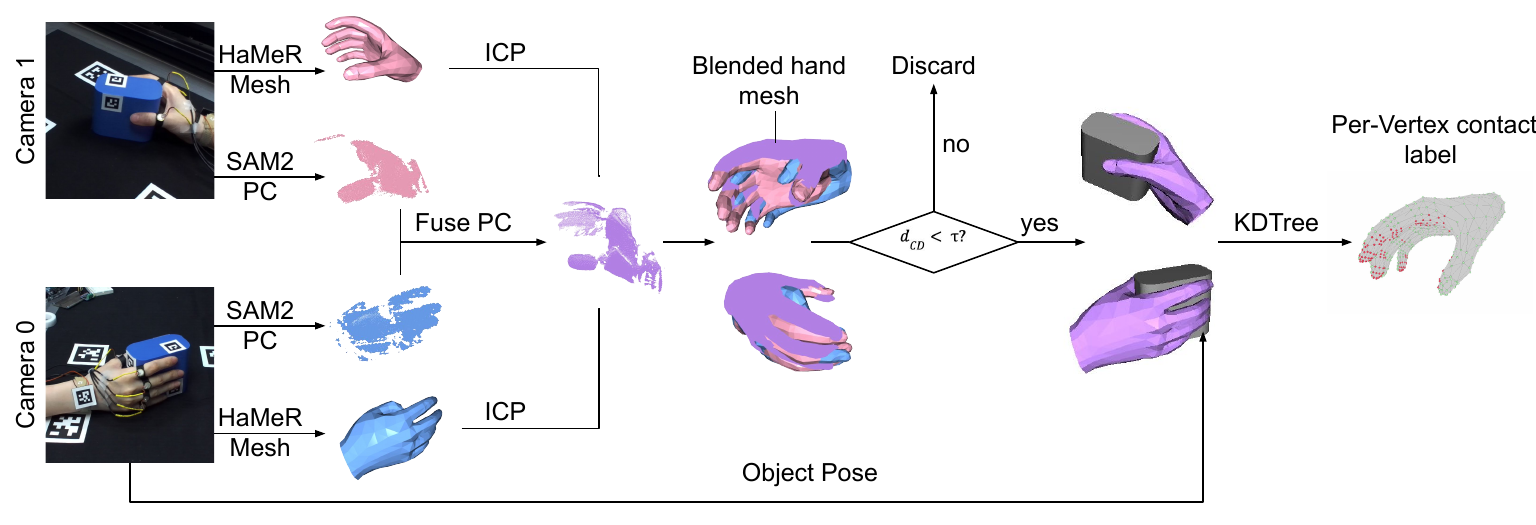}
\caption{\textbf{\algname ground-truth hand pose and contact annotation pipeline.} Our multi-view dataset collection approach integrates data from two RGB-D cameras to generate accurate hand meshes and contact annotations. Starting with RGB input from both cameras, we use SAM2~\cite{ravi2024sam2} for hand segmentation to extract point clouds (PC), while HaMeR~\cite{pavlakos2024reconstructing} generates initial MANO-based hand meshes. These components are fused through calibrated camera extrinsics, with ICP registration aligning the meshes to the combined point cloud. The blended mesh is validated against a chamfer distance threshold ($d_{CD}$), discarding frames with poor alignment. For valid frames, we track object pose using ArUco markers with ICP refinement, then we compute proximity between hand vertices and the object surface, generating per-vertex binary contact labels with a 5-mm threshold. This pipeline creates high-quality ground-truth data for training our visuo-acoustic model even in challenging contact-rich scenarios.}
\label{figure:contact_estimation}
\end{figure}

We take a vision-based approach to obtain ground-truth contact labels (Fig.~\ref{figure:contact_estimation}), addressing occlusion by fusing hand reconstructions from both camera views. For object tracking, we use ArUco markers with Iterative Closest Point (ICP) refinement to balance accuracy and efficiency.

\paragraph{Multi-view Hand Pose Fusion}
To reconstruct the hand pose, we integerate RGB-D from 2 complementary views with learned mesh priors. For each 60-s recording, we segment 3 keyframes per camera using SAM2~\cite{ravi2024sam2}, and propagate these masks to extract hand point clouds across time. In parallel, HaMeR~\cite{pavlakos2024reconstructing} estimates anatomically plausible hand meshes with 778 vertices using MANO~\cite{MANO2017}, which enforces consistent topology and incorporates learned kinematic priors. We first fuse the two hand point clouds using calibrated camera extrinsics and align each mesh to the fused point cloud via ICP. Then we merge the two ICP-aligned meshes using As-Rigid-As-Possible (ARAP) deformation, guided by the local geometry agreement. If the blended mesh fails to align with the fused point cloud, as measured by a chamfer distance threshold, we discard the corresponding frames to enhance label reliability. To ensure robust, stable tracking, we also mount an ArUco marker on the user’s wrist and perform an additional ICP alignment to that marker’s pose, constraining the hand mesh to a reasonable global location. This joint fusion gives an accurate, temporally coherent 3D reconstruction of the hand.

\paragraph{Object Tracking}
We attach an ArUco marker to each object and detect its 6-DoF pose relative to the camera in every frame. Using this initial estimation, we align the object mesh accordingly. To improve accuracy, the pose is further refined via ICP registration between the mesh and the depth data. This refinement yields accurate, frame-wise tracking of the object’s position and orientation.

\paragraph{Contact Estimation}
With hand and object meshes registered in a shared coordinate frame, we assign a binary contact label to each of hand mesh vertices based on proximity. A vertex is marked in contact if its nearest point on the object lies within a 5-mm threshold. This approach yields dense, frame-level contact annotations that are geometrically consistent and robust to occlusion.




\subsection{\algname Model}

The \algname architecture (Figure~\ref{figure:overview}) consists of three main components: an audio encoder, a mesh encoder, and a cross-modal fusion module for contact prediction.

\paragraph{Audio Encoder}

We process the continuous multi-channel acoustic signals to isolate and enhance contact-relevant features through the following steps: 
\begin{enumerate*}[label=(\emph{\roman*})]
\item \emph{reference subtraction}, where a baseline acoustic profile, captured during the first 5\,ms of each recording when the hand is free-floating without contact, is subtracted from subsequent frames to isolate contact-induced changes;
\item \emph{temporal alignment}, where the waveform is segmented into 35-ms windows corresponding to each video frame (30\,fps) with shifting windows to maintain synchronization between modalities; 
\item \emph{spectrogram extraction}, using a short-time Fourier transform (1024-point FFT, 512-point hop length) to yield time-frequency representations for all five microphone channels; and 
\item \emph{normalization}, applying frequency-bin normalization followed by per-channel standardization to ensure consistent feature representation and mitigate variations in microphone sensitivity introduced by various factors such as manufacturing variance.
\end{enumerate*}

This preprocessing yields aligned spectrograms that capture the characteristic spectral changes when the hand contacts objects. To extract discriminative features from these spectrograms, we use a pretrained VGG backbone that we finetune. The network processes each microphone channel independently before channel-wise feature fusion through self-attention, allowing the model to adaptively weight signals based on their relevance to contact events. The final audio embedding vector $\mathbf{z}_{\text{audio}} \in \mathbb{R}^{256}$ encodes the complex acoustic patterns resulting from both hand pose configuration and object contact interactions, providing complementary information to the visual modality.

\paragraph{Mesh Encoder}
The mesh encoder processes the hand geometry using a hierarchical graph neural network that preserves the mesh's topological structure. Given a hand mesh with $N=778$ vertices and an adjacency matrix derived from mesh connectivity, we apply a series of graph convolutional operations:
\begin{align}
\mathbf{H}^{(1)} &= \text{ReLU}(\text{GCN}(\mathbf{X}, \mathbf{A})) \\
\mathbf{H}^{(2)} &= \text{ReLU}(\text{GCN}(\mathbf{H}^{(1)}, \mathbf{A})) \\
\mathbf{H}^{(3)} &= \text{GAT}(\mathbf{H}^{(2)}, \mathbf{A}),
\end{align}
where $\mathbf{X} \in \mathbb{R}^{N \times 3}$ represents vertex coordinates, $\mathbf{A}$ is the adjacency matrix, GCN denotes graph-convolutional layers, and GAT represents a graph-attention layer with 4 attention heads. The resulting node embeddings $\mathbf{H}^{(3)} \in \mathbb{R}^{N \times 256}$ capture local geometric features for each vertex. In our notation, $\mathbf{H}^{(l)}$ represents the feature matrix for all vertices at the $l$-th layer of the graph neural network, while $\mathbf{h}_i^{(l)}$ denotes the feature vector for vertex $i$ at layer $l$. We compute a global mesh representation using a global attention pooling layer:
\begin{equation}
\mathbf{z}_\text{mesh} = \sum{i=1}^N \alpha_i \cdot \mathbf{h}^{(3)}.
\end{equation}
This hierarchical approach enables the model to capture both local vertex-level features and global hand configuration information. A key insight with this graph-based approach is that the hand's kinematics and use often creates long-range dependencies among contacts, where grasping typically engages opposing surfaces and are pose-dependent. The \algname architecture leverages this with progressively larger receptive fields in early layers followed by attention mechanism that can model anatomically far but functionally correlated contact regions. This information is pooled together while preserving relevant contact cues. This information is complemented by the acoustic modality to disambiguate contact conditions given the present pose of the hand.

\paragraph{Cross-Modal Fusion and Contact Prediction}
The fusion module combines audio and mesh representations to predict per-vertex contact probabilities. First, we concatenate global features from both modalities $\mathbf{z}_{\text{global}} = [\mathbf{z}_{\text{audio}}; \mathbf{z}_{\text{mesh}}] $.

This combined representation is processed through an MLP to extract cross-modal features:
\begin{equation}
\mathbf{z}_\text{fused} = f(\mathbf{z}_\text{global})
\end{equation}
We then compute per-vertex contact predictions by combining local vertex features with the global cross-modal representation:
\begin{align}
\mathbf{v}_i &= [\mathbf{h}_i^{(3)}; \mathbf{z}_\text{fused}] \\
\alpha_i &= \sigma(g_a(\mathbf{v}_i)) \\
\hat{y}_i &= \sigma(g_p(\mathbf{v}_i \odot \alpha_i)),
\end{align}
where $\mathbf{h}_i^{(3)}$ represents the feature vector for vertex $i$, $\alpha_i$ is an attention weight, and $\hat{y}_i$ is the predicted contact probability. This attention mechanism allows the model to focus on the hand's most relevant regions based on geometric and acoustic cues.

\subsubsection{Training Procedure}

We train our model using a weighted binary cross-entropy loss to address the significant class imbalance inherent in contact estimation, where contact vertices typically constitute only 5--10\,\% of the total vertices:
\begin{equation}
\mathcal{L} = -\frac{1}{N} \sum_{i=1}^{N} [w_1 y_i \log(\hat{y}_i) + w_0 (1-y_i) \log(1-\hat{y}_i)],
\end{equation}
where $y_i \in \{0,1\}$ is the ground-truth contact label for vertex $i$, and $w_1$ and $w_0$ are class weights for the positive (contact) and negative (non-contact) classes, respectively. We compute these weights inversely proportional to class frequencies in each training batch:
\begin{equation}
w_1 = \frac{N}{2 \cdot \sum_{i=1}^{N} y_i}, \quad w_0 = \frac{N}{2 \cdot \sum_{i=1}^{N} (1-y_i)}.
\end{equation}
This weighting scheme penalizes false negatives more heavily than false positives, encouraging the model to correctly identify the sparse contact regions despite their underrepresentation in the training data.

\subsection{Implementation Details}

\algname's mesh encoder has a hierarchical graph neural network with two GCN layers (64 and 128 channels) followed by a 4-headed graph attention layer (256 channels). We use the MANO~\cite{MANO2017} hand mesh topology to define edge connectivity for message passing operations. For training, we used the Adam optimizer with an initial learning rate of 0.001, batch size of 32 with gradient accumulation for effective batch sizes of 512, and a reduce-on-plateau scheduler with a factor of 0.5 and patience of 5 epochs. To mitigate overfitting, we utilize dropouts ($p=$ 0.2--0.3) throughout the network. The models were trained for 20 epochs on NVIDIA RTX 4090 GPUs. Each model took around 15 hours to train.

\section{Experiments}

We conducted an evaluation of \algname to assess its effectiveness in hand-object contact estimation across diverse scenarios. Our experiments were designed to validate key aspects: (1) the accuracy of the proposed visuo-acoustic approach compared to a vision-only baseline, (2) the robustness of the system under challenging conditions like occlusion and complex object geometry (3) the generalization capabilities across unseen objects and users.

We first describe our evaluation metrics to test both contact classification-based accuracy measures and geometric accuracy. We then present quantitative results comparing \algname to state-of-the-art vision-only baselines, followed by ablation studies that isolate the contribution of each modality and architectural component. Finally, we showcase qualitative results highlighting specific scenarios where our approach excels, particularly in cases where vision alone struggles to accurately determine contact regions.

\subsection{Evaluation Metrics}
We evaluate contact estimation with two complementary metrics: label accuracy and geometric precision. Both provide insight into how well \algname may work in comparison to state-of-the-art baselines. For label accuracy, we use the F1 score, defined as:
\begin{equation}
F_1 = \frac{2 \cdot P \cdot R}{P + R},
\end{equation}
where $P$, precision, represents the fraction of predicted contacts that are true: $P = \frac{|\mathcal{C}_{\text{pred}} \cap \mathcal{C}_{\text{true}}|}{|\mathcal{C}_{\text{pred}}|}$, and $R$, recall, quantifies the fraction of true contacts that are predicted: $R = \frac{|\mathcal{C}_{\text{pred}} \cap \mathcal{C}_{\text{true}}|}{|\mathcal{C}_{\text{true}}|}$. This metric effectively penalizes both false positives and false negatives in contact estimation, and a higher F1 score indicates better performance. 

For geometric precision, we compute the chamfer distance ($d_{CD}$) between the set of predicted contact vertices $\mathcal{V}_{\text{pred}}$ and the set of ground-truth contact vertices $\mathcal{V}_{\text{true}}$:
\begin{equation}
d_{CD}(\mathcal{V}_{\text{pred}}, \mathcal{V}_{\text{true}}) = \frac{1}{2|\mathcal{V}_{\text{pred}}|}\sum_{x \in \mathcal{V}_{\text{pred}}}\min_{y \in \mathcal{V}_{\text{true}}}\lVert x-y\rVert + \frac{1}{2|\mathcal{V}_{\text{true}}|}\sum_{y \in \mathcal{V}_{\text{true}}}\min_{x \in \mathcal{V}_{\text{pred}}}\lVert x-y\rVert.
\end{equation}
This metric averages the squared nearest-neighbor distances in both directions between the two sets, so a low chamfer distance indicates that predicted contact vertices lie very close in $\mathbb{R}^3$ to the true contacts. Together, the F1 score and chamfer distance ensure our evaluation captures both correct classification of contact regions and their localization with high spatial fidelity.

\begin{table}[t]
  \centering
  \caption{Performance comparison across different input modalities. We report average F1 Score ($\uparrow$) and average Chamfer Distance ($\downarrow$), along with performance on unseen objects and subjects.}
  \label{tab:modality_comparison}
  \resizebox{\textwidth}{!}{
  \begin{tabular}{llcccccc}
    \toprule
    \textbf{Method} & \textbf{Modality} 
    & \textbf{Avg F1 ($\uparrow$)} 
    & \textbf{Avg Chamfer ($\downarrow$)} 
    & \textbf{Unseen Obj F1 ($\uparrow$)} 
    & \textbf{Unseen Subj F1 ($\uparrow$)} 
    & \textbf{Unseen Obj Chamfer ($\downarrow$)} 
    & \textbf{Unseen Subj Chamfer ($\downarrow$)} \\
    \midrule
    \makecell{Hold~\cite{fan2024hold} \\ (Baseline)} 
    & RGB 
    &0.1171 $\pm$ 0.0974 &  31.33 $\pm$ 27.36\,mm & 0.1431 $\pm$ 0.1026 & 0.1325 $\pm$ 0.0889 & 46.55 $\pm$ 13.45\,mm & 19.92 $\pm$ 5.489\,mm \\
    
    \makecell{\algname \\ (Proposed)} 
    & RGB + Audio 
    &\textbf{0.327} $\pm$ 0.122 & \textbf{5.487} $\pm$ 1.612\,mm & \textbf{0.288} $\pm$ 0.111 & \textbf{0.302} $\pm$ 0.116 & \textbf{6.042} $\pm$ 1.755\,mm & \textbf{6.828} $\pm$ 1.704\,mm \\
    \bottomrule
  \end{tabular}
  }
\end{table}

\begin{figure}[t]
\centering
\includegraphics[width=1.0\columnwidth]{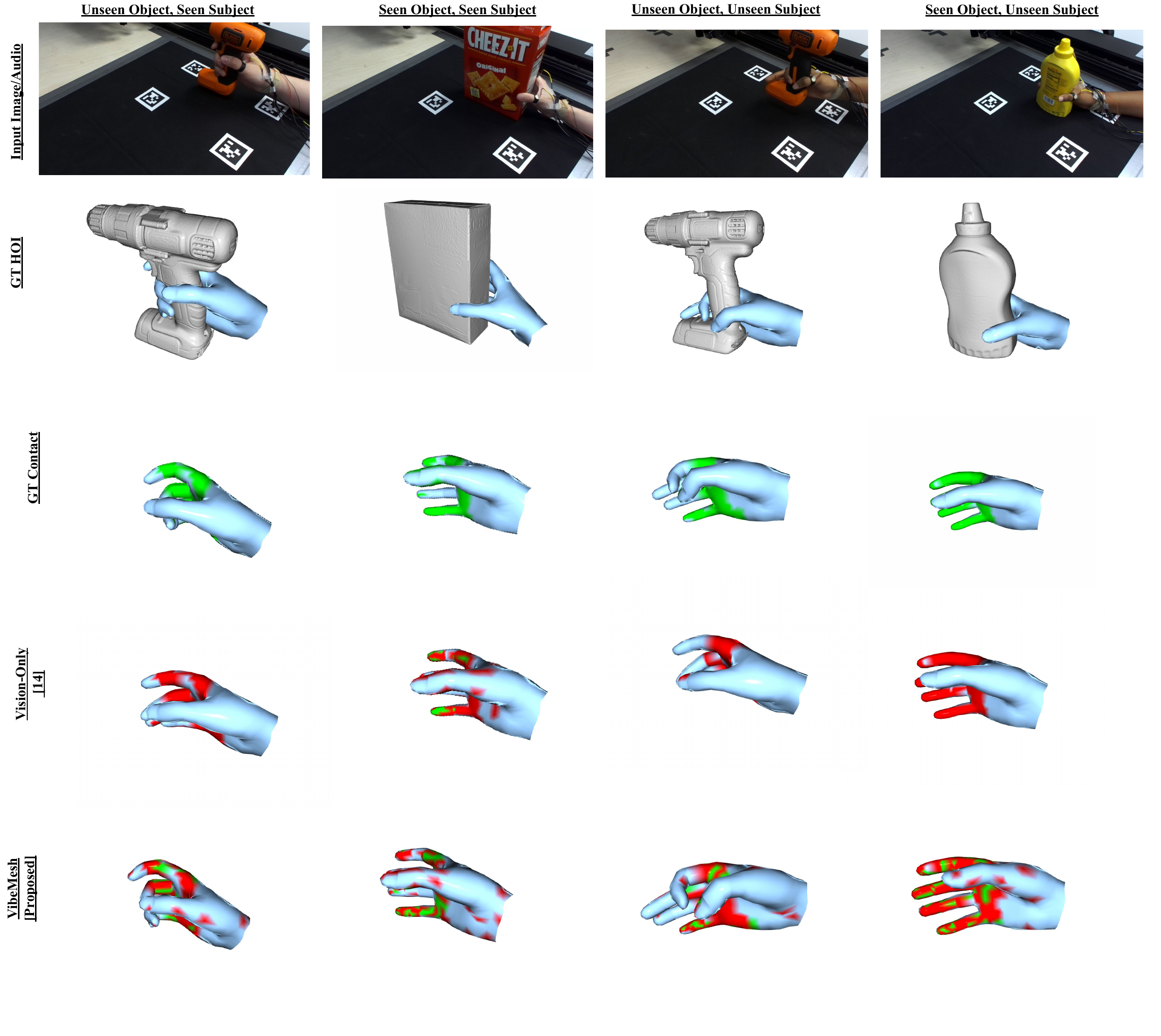}
\caption{\textbf{Qualitative Results}. Each column represents a different test condition: unseen object with seen subject, seen object with seen subject, unseen object with unseen subject, and seen object with unseen subject. The rows show: (1) input RGB images to the models with the subjects interacting with objects while wearing the active acoustic sensing platform (2) ground-truth hand-object interaction (GT HOI) visualizations showing 3D hand meshes and object models (3) ground-truth contact labels with contact vertices highlighted in green (4) contact prediction results from the vision-only baseline~\cite{fan2024hold}, showing substantial false negatives (missing contacts) and false positives (incorrect contact regions) in red (5) \algname predictions, demonstrating noticeable improvements in contact localization across all generalization scenarios. We note that \algname largely identifies contact points even in challenging cases with partial occlusion and novel objects/subjects. }
\label{figure:task_results}
\vspace{-0.7 cm}
\end{figure}

\subsection{Results} Our experiments demonstrate that \algname noticeably outperforms vision-only and audio-only approaches across various scenarios. Table~\ref{tab:modality_comparison} shows that \algname achieves an average F1 score of 0.327, representing a 179.4\% improvement over the state-of-the-art vision-only Hold~\cite{fan2024hold} baseline (F1 score of 0.117). The substantial performance gap is evident in the chamfer distance metric, where \algname (5.487\,mm) achieves an 82.5\% reduction compared to the vision-only baseline (31.33\,mm). This supports the complementary nature of the visuo-acoustic approach, where each modality contributes unique information about contact conditions, which we study further in ablation studies in the later sections.

When analyzing performance on the challenging case of unseen objects, \algname demonstrates robust generalization capabilities with an F1 score of 0.288, significantly outperforming the vision-only baseline (0.143) by 101.4\%. This suggests that our cross-modal approach captures generalizable features of hand-object interactions that transfer effectively to novel objects of different geometries and materials. 

Similarly, the chamfer distance on unseen objects (6.042\,mm) shows an 87.0\% improvement over the vision-only approach (46.55\,mm), indicating that our method accurately localizes contact points even on previously unseen geometries. Cross-user evaluation further highlights \algname's generalization abilities across different hand sizes and interaction styles, maintaining an average F1 score of 0.302. This represents a 7.6\% decrease from the overall average, indicating that the learned visuo-acoustic features capture fundamental patterns of hand-object contact that transcend individual differences in hand geometry and manipulation behavior. 

The cross-user chamfer distance (6.828\,mm) shows a 65.7\% improvement over the vision-only baseline (19.92\,mm), demonstrating that our approach maintains spatial precision even when applied to previously unseen hand geometries. Qualitative results in Figure~\ref{figure:task_results} illustrate several challenging scenarios where \algname excels. Under partial occlusion of object geometry and hand, where the vision-only baseline struggles to accurately determine contact regions, \algname’s visuo-acoustic approach correctly identifies contact regions by leveraging the acoustic signals that propagate through the hand. 
\subsection{Baselines and Ablation}

We conduct an ablation study to examine the contribution of each component in the \algname architecture, with results presented in Table~\ref{tab:ablation}. Removing the audio modality (``w/o Audio'') results in a 70.6\% decrease in F1 score and an 80.0\% increase in chamfer distance, confirming that acoustic signals provide critical information for contact estimation that vision alone cannot capture. This degradation is most pronounced during contact-rich cases, where visual cues become less reliable.

Conversely, the ``w/o Vision'' variant—which relies solely on acoustic features—shows a 33.3\% reduction in F1 score compared to the full model. While acoustic sensing excels at detecting transient events and materials properties, it lacks the spatial precision that visual information provides, particularly for localizing contacts on the hand mesh. This underscores the complementary nature of the two modalities.

The ``w/o Fusion Module'' variant, which processes audio and visual features independently before simple concatenation, shows a 12.2\% decrease in F1 score. This highlights the importance of our cross-modal attention mechanism for integrating information from both modalities. The attention-based fusion learns modality-specific reliability, focusing on acoustic cues when vision is unreliable and leveraging visual precision when available.


\begin{table}[t]
  \centering
  \caption{Performance comparison across different input modalities. We report average F1 Score ($\uparrow$) and average Chamfer Distance ($\downarrow$), along with performance on unseen objects and subjects.}
  \label{tab:modality_comparison}
  \resizebox{\textwidth}{!}{
  \begin{tabular}{lllcccccc}
    \toprule
    \textbf{Method} & \textbf{Modality} 
    & \textbf{Avg F1 ($\uparrow$)} 
    & \textbf{Avg Chamfer ($\downarrow$)} 
    & \textbf{Unseen Obj F1 ($\uparrow$)} 
    & \textbf{Unseen Subj F1 ($\uparrow$)} 
    & \textbf{Unseen Obj Chamfer ($\downarrow$)} 
    & \textbf{Unseen Subj Chamfer ($\downarrow$)} \\
    \midrule
    w/o Audio & RGB & 0.096 $\pm$ 0.071 & 9.853 $\pm$ 2.140 & 0.078 $\pm$ 0.062 & 0.082 $\pm$ 0.060 & 10.215 $\pm$ 2.356 & 9.934 $\pm$ 2.218 \\
    w/o Vision & Audio & 0.218 $\pm$ 0.103 & 7.642 $\pm$ 1.872 & 0.186 $\pm$ 0.092 & 0.195 $\pm$ 0.098 & 8.124 $\pm$ 2.065 & 7.985 $\pm$ 1.943 \\
    w/o Fusion Module & RGB + Audio & 0.287 $\pm$ 0.115 & 6.319 $\pm$ 1.743 & 0.251 $\pm$ 0.108 & 0.265 $\pm$ 0.112 & 6.875\,mm $\pm$ 1.826 & 6.742 $\pm$ 1.798\,mm \\
    \makecell{\algname } 
    & RGB + Audio 
    &\textbf{0.327} $\pm$ 0.122 & \textbf{5.487} $\pm$ 1.612\,mm & \textbf{0.288} $\pm$ 0.111 & \textbf{0.302} $\pm$ 0.116 & \textbf{6.042} $\pm$ 1.755\,mm & \textbf{6.828} $\pm$ 1.704\,mm \\
    \bottomrule
  \end{tabular}
  }
  \label{tab:ablation}
\end{table}

\section{Conclusion}
We present \algname, a novel visuo-acoustic approach for hand pose and contact estimation that integrates lightweight wearable acoustic sensors with visual observation. Our results demonstrate that this multi-modal system significantly outperforms vision-only approaches, particularly in challenging scenarios involving occlusions and static contacts. By fusing complementary information from acoustic propagation patterns and visual hand reconstruction, \algname achieves more robust and accurate contact estimation across diverse objects and users.

\algname's key insight to the task is that hand-object contacts modify the acoustic transmission properties of the hand in ways that can be measured and interpreted, even when vision may face contact condition ambiguities and fail to capture these interactions. \algname's graph-based attention network effectively integrates these cross-modal signals, learning to leverage the strengths of each modality while compensating for their individual limitations. The result is a contact estimation system that maintains high performance across varied conditions.

\section{Limitations}
While \algname demonstrates consistent improvements over vision-only approaches, it has limitations. First, although the piezoelectric contact microphones mechanically reject ambient noise by attending to only solid contact-based sound transmission, we did not include results on \algname's robustness to exceptionally loud environments. However, we note that although common ambient sounds such as verbal conversations were present during data collection, the microphone signals were largely unaffected. Second, we relied on multi-view camera systems with fiducial markers for perceiving hand poses and hand-object interactions, which are inherently indirect methods of observing contacts. Although there are some commercially available force or contact sensorized gloves that could enable us to collect hand-object interactions directly, they are expensive and suffer from low spatial resolution. Finally, \algname largely treats pose and contact estimation in a sequential manner, where we first estimate pose with a visual model and estimate contact conditioned on both the pose estimation and acoustic signals. In the future, we may improve on the approach by integrating both vision and acoustic signals to simultaneously reason about pose and contact to benefit from both modalities at each step. 

\bibliographystyle{plainnat}
\bibliography{ref.bib}

\end{document}